# A wave prediction system for real time sea state forecasting in Black Sea

Anna Kortcheva,  Marieta Dimitrova,  Vasko Galabov

*  National Institute of Meteorology and Hydrology, 66 Tzarigradsko shausse, 1784 Sofia, Bulgaria

## Abstract

This paper briefly describes the existing operational system for wind waves forecasting in the Black Sea.  It is a system of coupled atmospheric and wave numerical models aiming at a detailed and accurate sea state forecast on an operational level. The system was created at the National Institute of Meteorology and Hydrology Bulgarian Academy of Sciences (NIMH-BAS) in collaboration with the Meteorological Office of France - Meteo-France. The present work introduces the use of wave models at NIMH-BAS and shows the model results, as well as an intercomparison. The numerical wave models VAG, WAVEWATCH III and WAM, developed by the research groups of Meteo-France, NCEP and WAMDI, have been adopted for the Black Sea area and implemented at the NIMH-BAS to allow real-time forecasts and hindcasts of the waves in the Black Sea. The coupling of two atmospherics models ARPEGE and ALADIN has been used to force the wave models. The operational use has indicated that the system is suitable for general purposes and the results are generally satisfactory. The wave models are evaluated through a comparison with the altimeter satellite measurements. A comparison between the model results and the altimeter data from the satellites ERS1/2 and ENVISAT demonstrates that the models call fairly reproduces the observed characteristics of waves.

**1. Introduction.**
Wind waves affect human activities and have great geophysical significance. The description of the sea-state now is mandatory in the international safety bulletins. Accurate forecasting of sea conditions and adequate forewarning is essential for the safety of people, equipment and the environment. One other important need is related to the management of activities at the coastal zone by providing climatological data, hindcasted by the numerical wave models.
The sea state forecasting system operated by NIMH-BAS consists of the numerical weather prediction (NWP) models ARPEGE and ALADIN and of tree spectral wave models -VAGBUL (in operational use), WAVEWACH III referred to as WW3 (in operational use) and WAM (non-operational).

The operational sea waves predictions of NIMH-BAS are performed using the wave model VAGBUL and the operational products of the atmospheric numerical weather prediction (NWP) models of Meteo-France - ARPEGE and ALADIN, as an input. This combination of a wind input and a wave model is called the Black Sea wind wave forecasting system.Alternately, a third generation WAM model cycle, developed by WAMDI-group (1988), and WAVEWATCH III version 2.22, developed by NCEPT (Tolman, 2002), have been introduced to NIMH-BAS since 2003. Using the same domain and spatial resolution as VAGBUL, the models are ruining in a quasi-operational manner and are used for intercomparison studies.

The wind wave prediction system was created in 1996 as a result of the common efforts of the scientists of NIMH-BAS and Meteo-France. This system is the part of the Marine forecasting system of NIMH-BAS (Fig.1).

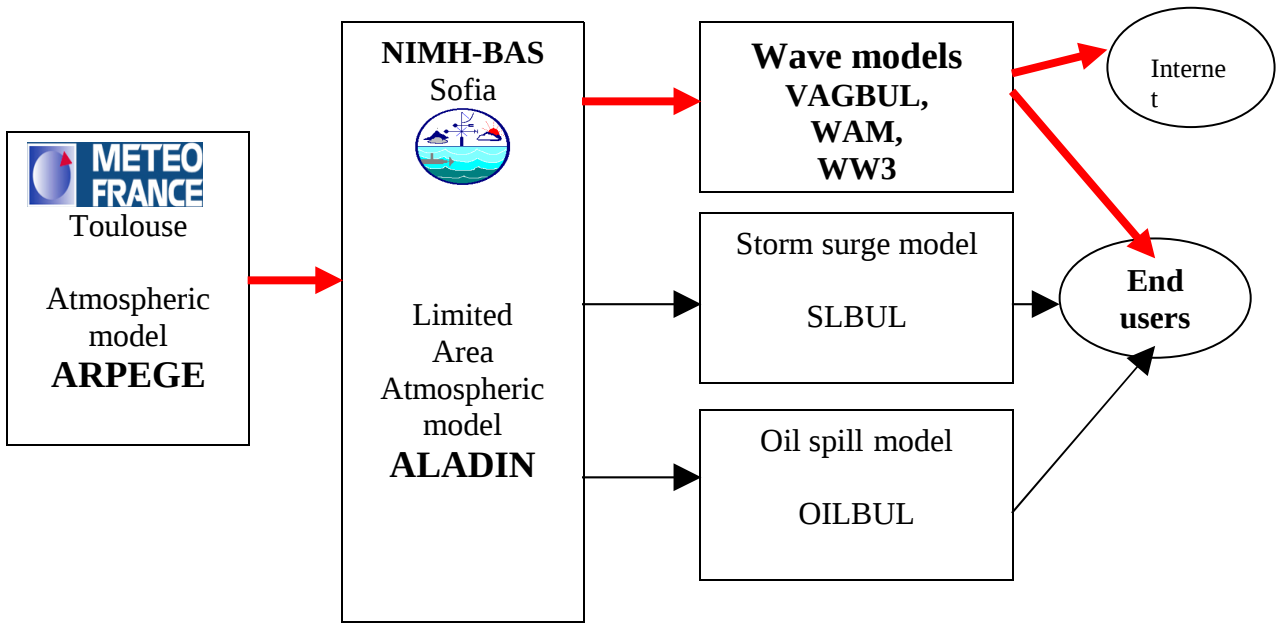

Fig.1. Marine forecasting system of NIMH-BAS

In section 2 we briefly introduce the VAGBUL, WW3 and WAM wave models. A comparison between model results and satellite altimeter measurements, as well as a model intercomparison, are shown in section 3. Section 4 gives a summary of the conclusions and an outlook on future works.

## 2. Wave models

NIMH-BAS operationally uses numerical wave models, which provide detailed sea-state information for any given location in the Black sea area. This information is modelled in the form of two-dimensional (frequency-direction) spectra.

NIMH-BAS operates tree models for ocean wave prediction. One is the second-generation wave model VAGBUL and the other two are the third- generation wave models WW3 and WAM. The current VAGBUL has been in operational use since March 1996, WW3 and WAM since January 2003.

The specifications of the wave models are summarized in Table 1. The numerical models VAGBUL, WAM and WW3 have been implemented on a spherical grid cover the area of the Black Sea from 40°N to 47°N and from 27°E to 42°E on a regular latitude-longitude grid. The grid resolution is 0.25°x0.25°. The geographic coverage of the models and the numerical grid are shown in Fig. 2.

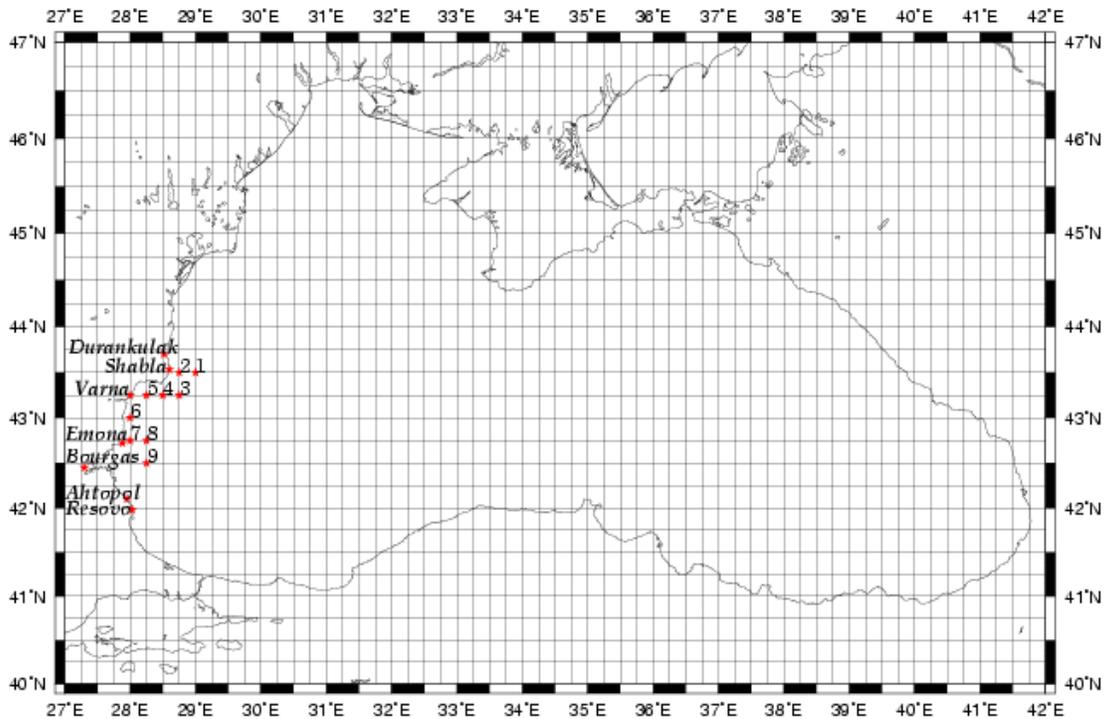

Fig.2. Model domain used at NIMH-BAS. This map also shows the grid points (1-9) located in the western part of the Black Sea, at which the intercomparison between the VAGBUL (two versions) and WW3 wave models results is systematically performed.

Computer resources
The wave models have been implemented on a PC with two processors, operating under Linux.

Graphics software
The Grid Analysis and Display System (GrADS) has been used for the plot of the WW3 results. The Generic Mapping Tools (GMT) have been used for the output maps of the WAM and VAGBUL wave models.

| MODEL | VAGBUL second generation | WW3 (version 2.22) third generation | WAM (cycle 4) third generation |
|---|---|---|---|
| Domain | Black Sea area | | |
| GRID | 0.25 x 0.25 degree latitude/longitude spherical grid 61 x 29 grid points | | |
| Spectral discretization | 22 frequency bands 18 directional bands | 25 frequency bands 24 directional bands | 25 frequency bands 24 directional bands |
| First/last frequency | 0.040 /0.296 Hz | 0.042 /0.411 Hz | 0.042 /0.411Hz |
| Start of forecast T | 00 and 12 UTC | | hindcast |
| Forecast range | T+48 | | hindcast |

Table 1. Main characteristics of the wave models

*VAGBUL*

The VAG model, developed at Meteo-France (Guillaume, 1987; Fradon et al, 2000; Lefevre et al, 2003) has been operational at NIMH-BAS since 1996 as a version VAGBUL. VAG performance was improved by introducing a new physics package proposed by Fradon (1997) and implemented in the last version of the VAG model by Stefanescu (2001). The purpose of these modifications was to tend to a better balance between growth and decay when the waves are fully developed. This was partly achieved by using the same expressions in VAG to describe growth and decay as the ones used in WAM. The modified version of VAG has been implemented in NIMH-BAS in 2006, as a version VAGNEW, but the previous version of VAGBUL (VAGOLD) still remains in operational use.
VAGBUL has been formulated in terms of the basic transport equation for two-dimensional wave spectrum. The evolution of the two-dimensional ocean wave spectrum $E(f,\theta,\varphi,\lambda,t)$ with respect to the frequency f and the direction θ as a function of the latitude φ and the longitude λ on a spherical earth is governed by the transport equation:

$$\frac{\partial E}{\partial t} + \frac{1}{\cos\varphi}\frac{\partial}{\partial\varphi}(\frac{d\varphi}{dt}\cdot\cos\varphi\cdot E) + \frac{\partial}{\partial\lambda}(\frac{d\lambda}{dt}E) + \frac{\partial}{\partial\theta}(\frac{d\theta}{dt}E) = S \quad (2.1)$$

where: S is the net source function describing the wind input Sin, Sds is the surface dissipation, Snl is the non-linear interaction term and Sbf is the bottom friction.

The left-hand side of the energy balance equation (2.1) represents the propagation of the wind waves, i.e., the advection The first term of is presents the local variation of the E, terms 2 and.3 are present the latitude longitude propagation and term 4 the change of the wave direction by propagation to the spherical geometry.

$$\frac{d\varphi}{dt} = c_g R^{-1}\cos\theta \quad (2.2)$$

$$\frac{d\lambda}{dt} = c_g \sin\theta(R\cos\varphi)^{-1} \quad (2.3)$$

$$\frac{d\theta}{dt} = c_g \sin\theta\tan\varphi R^{-1} + \frac{1}{kR}\frac{\partial\omega}{\partial h}(\sin\theta\frac{\partial h}{\partial\varphi} - \frac{\cos\theta}{\cos\varphi}\frac{\partial h}{\partial\lambda}) \quad (2.4)$$

where g is the acceleration of gravity
R is the radius of the earth
h is the depth
k is the wavenumber

$$\omega = (gk\tanh kh)^{1/2} \quad (2.5)$$

The great circle refraction term for the deep water was augmented to include the refraction due to the variations of the water depth (Phillips,1977) - the second term in (2.4). The infinite group velocity $c_g$ for the deep water was replaced by the corresponding expression for finite depth h:

$$c_g = \frac{c}{2}(1 + \frac{2kh}{\sinh(2kh)}) \quad (2.6)$$

where c is the phase velocity

$$c = \sqrt{\frac{g}{k}\text{th}(kh)} \quad (2.7)$$

the source functions :

$$\frac{\partial E}{\partial t} = S = S_{in} + S_{ds} + S_{nl} + S_{bf} \qquad (2.8)$$

An important aspect of the wave model is its modular structure. Namely, two main tasks: integration for the advection term and for the source term are isolated. A first order "upstream" propagation scheme was implemented for the integration of the advection term. For the integration of the source term, a first order explicit scheme was implemented.

## *WAM*

NIMH-BAS also uses the WAM (cycle 4) wave model developed by the WAMDI-group (1988) for retrospective simulations. The WAM (WAMDI Group 1988) wave model is one of the best-tested wave models in the world. It has been distributed among over 50 research groups for forecasting on global and regional scales. WAM is operational at the European Centrum for Medium Range Weather Forecast (ECMWF). The wave model WAM solves explicitly (without any assumptions on the shape of the wave spectrum) the same energy balance equation used in the VAG model, but with additional non linear inter-actions terms. The physical part of the WAM model differs quite appreciably from that of the original VAG model. The source terms and propagation are computed with different methods and time steps. The source terms integration is done with an implicit integration scheme, while the propagation scheme is a first order upwind flux scheme.

## *WAVEWATCH III (WW3)*

Alternately to the two versions of VAGBUL, a third generation WW3 version 2-22 developed by NCEPT (Tolman, 2002) was introduced to NIMH-BAS in 2003. The third-generation wave model WW3 utilizes parallel programming and has different wave physics and a different numerical scheme from the VAGBUL wave model. Using the same domain and spatial resolution as VAGBUL, WW3 is ruining in a quasi-operational manner. WW3 behavior has been studied and compared with that of the VAGBUL model, which is simpler and uses less computing time.
 WW3 solves the spectral action density balance equation for wave number-direction spectra $N(k, \theta, \lambda, \varphi, t)$:

$$\frac{\partial N}{\partial t} + \frac{1}{\cos\phi} \frac{\partial}{\partial \phi} \dot\phi N \cos\theta + \frac{\partial}{\partial \lambda} \dot\lambda N + \frac{\partial}{\partial \theta} \dot\theta_g N = \frac{S}{\sigma}, \qquad (2.9)$$

## Wave models output

Wave energy two-dimensional spectra are post processed to provide output fields of the following variables: significant wave height (SWH), mean wave direction, mean period. An example of VAGBUL and WW3 wave output can be seen in Figures 3 and 4.

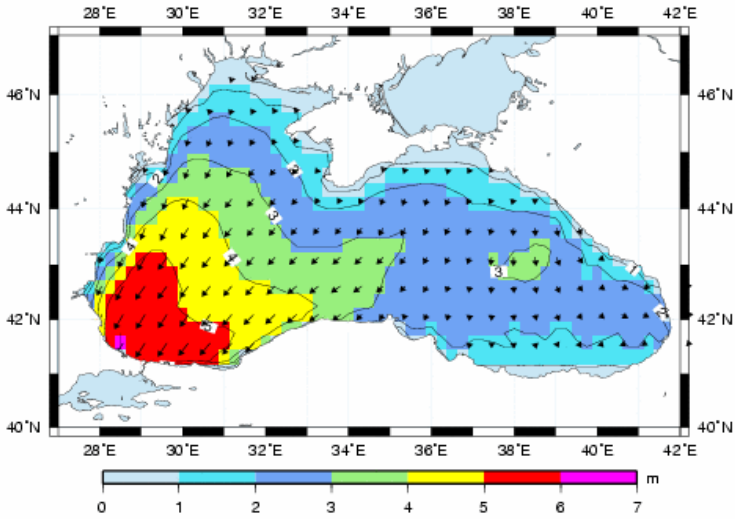

Fig. 3  Significant wave height and mean wave direction (unit vectors) on 24.01.2006 at 00.00 UTC computed by the VAGBUL model.

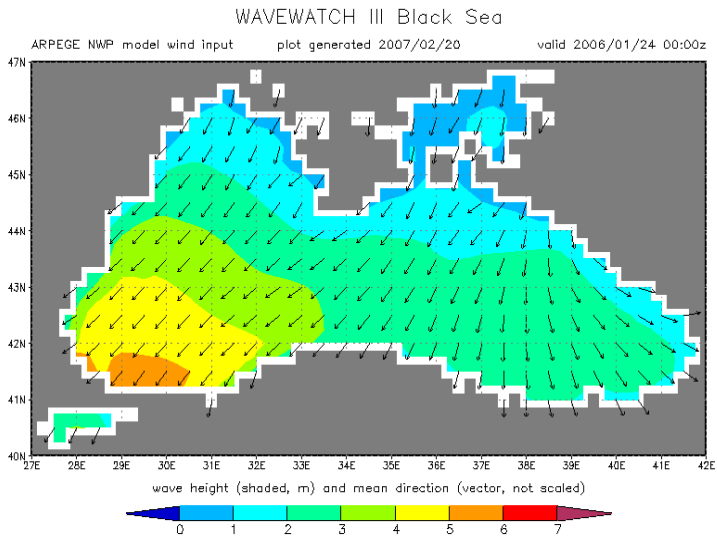

Fig. 4.  Significant wave height and mean wave direction (unit vectors) on 24.01.2006 at 00.00 UTC computed by the WW3 model.

Referring to Fig. 3 and 4 one can make a visual comparison of the output of both models. The maximum wave height calculated from the VAGBUL model is a little larger than that gained from the WW3 model

## Wind forcing

Wind fields for VAGBUL and WW3 wave models are obtained by coupling the NWP models ARPEGE and ALADIN. ARPEGE is the acronym of "Action de Recherche Petite Echelle Grande Echelle". This model is used both by METEO-FRANCE and ECMWF; at ECMWF it is named IFS (Integrated Forecasting System). ARPEGE is a global spectral model, with a Gaussian grid for the grid-point calculations. ALADIN is a limited-area version of ARPEGE. It is still a spectral model; its horizontal domain covers only a limited area, so the fields are ``bi-periodicised'' in order to match the spectral representation. ALADIN needs to be forced by a global model – ARPEGE, which has to provide lateral boundary conditions. The version of the ALADIN model, currently used as operational in NIMH-BAS, is CY29T2OP2 (Bogatchev A., 2006). Thus, if we say that the wave models are forced with ARPEGE fields, this is correct only when the coupling files are ARPEGE production. The obtaining of forcing fields is done using the ALADIN model on the grid of the coupling domain.

The wind fields are available at 6-hour intervals to (T+48) on a regular latitude-longitude grid with a 0.25°x025° mesh size. Fig.5 shows the wind field from AREPEGE during the winter storm from January 2006.

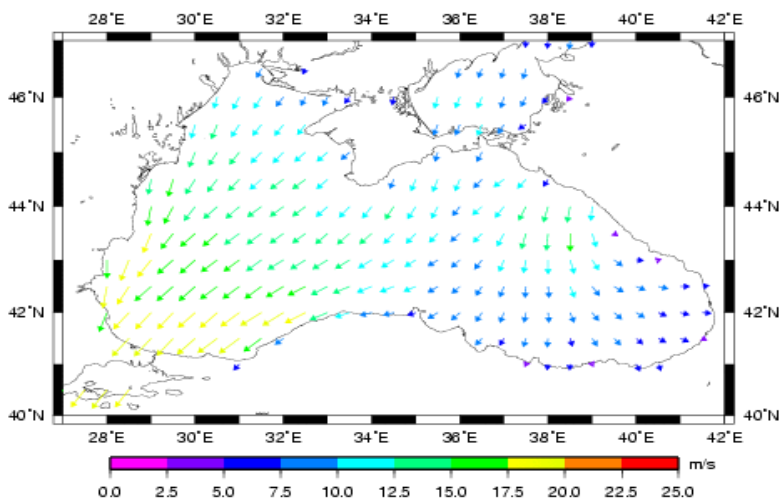

Fig. 5.  Wind field provided by  ARPEGE/ALADIN  NWP models
       on 24.01.2006 at 00 UTC (24 hour forecast from 23.01.2006)

3. Verification

The quality of the significant wave height (SWH) obtained from the VAG model, using the wind forcing from the ARPEGE NWP model has been assessed during the last few years. Excellent results were obtained in Meteo-France in the comparison study of the VAG, WAM and WAVEWATCH III wave models. (Lefevre, Kortcheva, Stefanescu, 2003).

A general problem for the verification of wave models over the Black Sea is the lack of wave data from buoys. Recent advances in satellite technology have created a possibility to utilise remotely sensed wave and wind data. From radar altimeter data one can estimate significant wave height and wind speed on the sea surface. Comparisons of the altimeter measured wind speed and wave height with surface buoy data have shown very positive agreement. With a resolution of 7 km, the spatial resolution of the space borne altimeter is comparable with that of the numerical models used for regional simulations.

The significant wave heights provided by the ERS1, ERS2 and ENVISAT satellites have been used for comparisons with VAGBUL and WW3 outputs in order to assess the hindcasts of the models during the storm situations over the Black Sea.

In the following, we present the results of a comparison study of VAGBUL and WW3 wave modeling over the Black Sea with the altimeter data from ERS1/2 and ENVISAT satellites.

Comparison of SWH modeled by VAGBUL against ERS1/2 measured SWH.

An evaluation of the VAGBUL wave model for the Black sea area and a comparison with data from the First and Second European Remote Sensing Satellites (ERS1, ERS2) has been carried out at Meteo-France for 15 severe storms registered from November 1994 to March 1998. (Kortcheva, 1997). Comparisons of the VAGBUL model derived SWH were made with the available ERS1/2 data. The satellite altimeter data have been collocated with the grid points of the VAGBUL wave model. Several statistical parameters have been calculated at each of the wave model's locations, collocated with the satellite track, and magnitude and variation of these parameters has been analyzed to determine the skill of the operational wave model. A total number of 687 collocated data points were selected over the period of 1994–1997. The overall statistics for SWH obtained from the VAGBUL forced by ARPEGE wind are presented in Table 2.

| N=687 | MEAN (m) | BIAS (m) | Standard Deviation | RMSE | Scatter index |
|---|---|---|---|---|---|
| VAGBUL SWH | 1.88 | 0.11 | 0.51 | 0.53 | 0.29 |
| ERS2 SWH | 1.78 | | | | |

Table 2. Verification statistics VAGBUL SWH against ERS1/2 altimeter data

The main indicator of the skill of the model is the SWH *RMS* error. It can be seen that VAGBUL wave model results show a good agreement with ERS1/2 satellite observations with RMSE equal to 0.53 m.

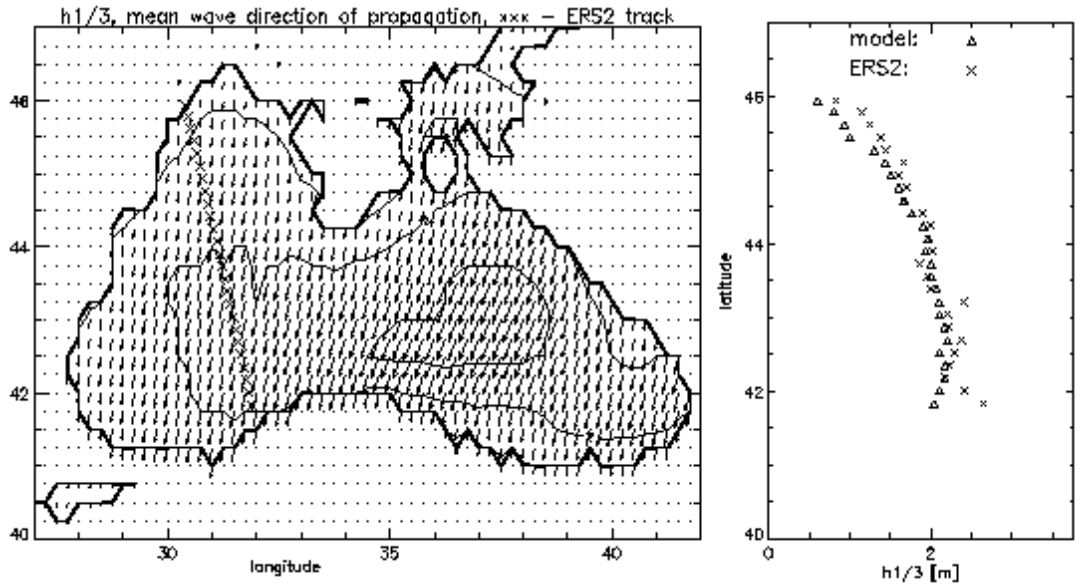

Fig. 6. Comparison of modeled by VAGBUL SWH and ERS-2 altimeter data during the storm in the Black Sea on 23 December 1996 at 12.00 UTC (A. Kortcheva, 1997).

Fig.6 presents the comparison between the modeled and the satellite significant wave height. It also illustrates the good agreement between the model and observed SWH along the ERS2 satellite track.

## Comparison of SWH modeled by VAGBUL and WW3 against ENVISAT measured SWH

The current validation of the atmospheric and the wave models is performed against altimeter data from the satellites JASON and ENVISAT. In order to investigate the ability of the different models, both VAGBUL and WW3 wave models have been employed to simulate wind waves induced by 10 severe storms in the Black Sea for the period 2003-2007, and the model results were compared with altimeter observations from the ENVISAT satellite. This study is still under development. The preliminarily results are shown below.
The evaluation is presented in the form of scatter plots of model SWH versus ENVISAT measured values (Figs.7 and 8); also verification statistics, such as mean error, root-mean-square error, etc. are presented in Table 3.

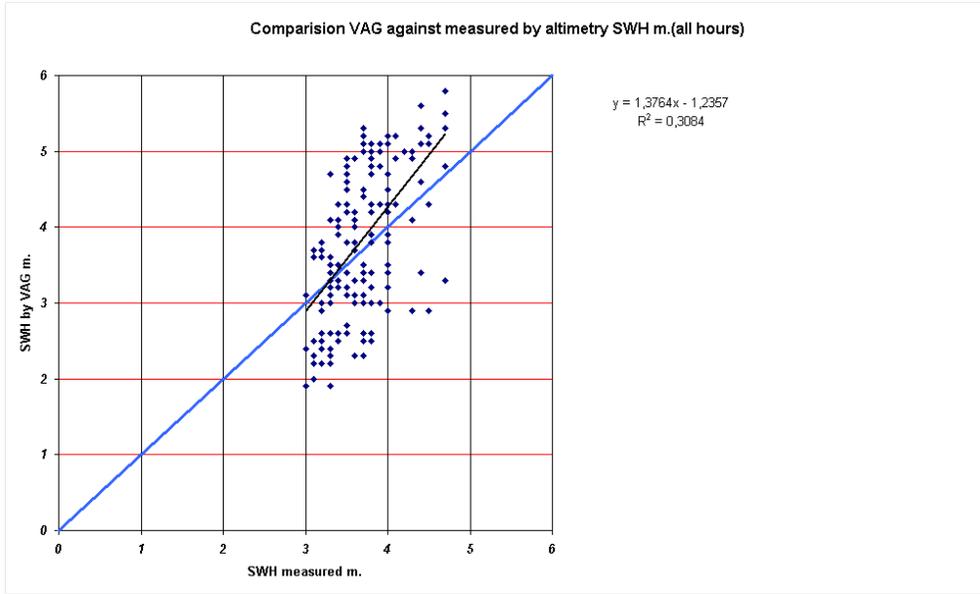

Fig. 7. Scatter plot for SWH from the VAGBUL wave model
against ENVISAT altimeter wave data (Galabov V, 2006)

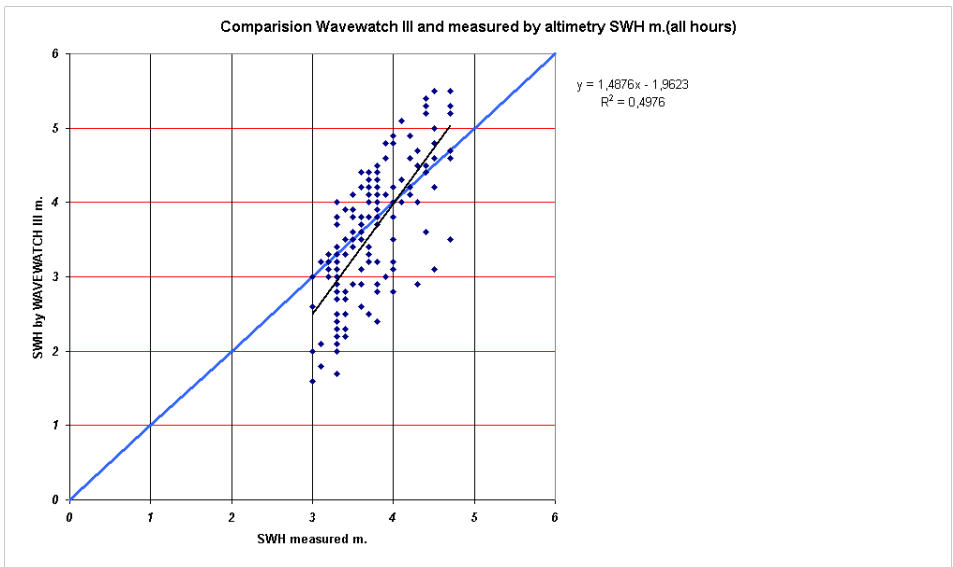

Fig. 8 Scatter plot for SWH from WW3 wave model
against ENVISAT altimeter wave data (Galabov V, 2006)

| SWH (m) | BIAS | STANDARD Deviation | RMSE | Scatter index | Symmetric slope |
|---|---|---|---|---|---|
| **VAGBUL** | 0.48 | 0.62 | 0.78 | 0.17 | 1,14 |
| **WW3** | 0.12 | 0.47 | 0.49 | 0.12 | 1.04 |

Table 3. Verification statistics of the VAGBUL and WW3 models' result against ENVISAT altimeter data (N=378) (SWH >3m)

The preliminary statistical results show that WW3 results are better than these of VAGBUL in comparison with ENVISAT altimeter data. This comparison study will be continued.

## Intercomparison VAGBUL (two versions) and WW3.

The VAGBUL model output in terms of SWH and mean period for selected grid points (Fig.2) is systematically evaluated against the WW3 model results over the Black Sea. The time series of SWH and the mean period for the period 23-25 January 2006 are plotted in Fig. 9.

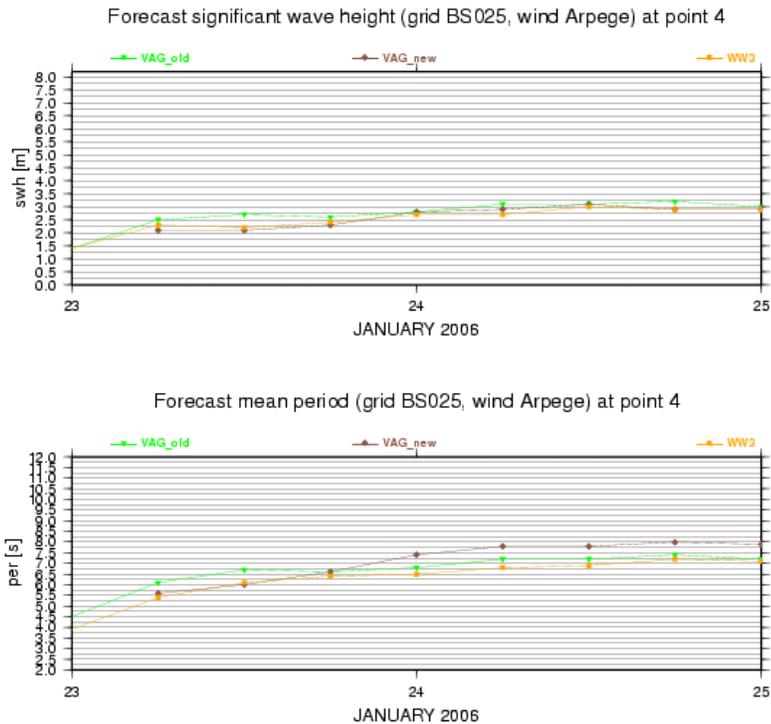

Fig.9. The time series of the SWH (top panel) and the mean wave period (bottom panel) from VAGBUL(VAGold), VAGnew and WW3 models at the location N4 from 23 to 25 January 2006 (Dimitrova M, Kortcheva A, 2006)

One can see a good agreement between the results of the models at the location N4. Unfortunately, there are no buoy observations available at the NIMH-BAS during the storm situations.

## 4. Conclusions and future work

The most important output parameter of the wave models, namely the significant wave height, has been evaluated against altimeter satellite data from ERS1/2 and ENVISAT. The evaluation is presented in the form of scatterplots of model versus altimeter values; also verification statistics such as mean error, root-mean-square error, etc. are presented for the two comparison studies.

The verification results suggest that the present operational version of NIMH-BAS can provide wave height forecast and hindcast with a considerable accuracy. The main indicator of the skill of the model is the SWH *RMS* error. It can be seen that the VAGBUL wave model results show a good agreement with ERS1/2 satellite observation with RMSE equal to 0.53 m.

If one considers the satellite observations from ENVISAT as a reference, the WW3 SWH is slightly better than the results from VAGBUL. When considering the comparison between models and ENVISAT satellites data, WW3 presents less scatter and lower errors for SWH than VAGBUL.

This study shows that the second-generation VAG model is nearly as good in predicting wave heights as the third-generation model WW3. Although the results presented here are encouraging, more observational, especially buoy data are highly needed for the model verification.

The NINH-BAS staff will continue to monitor the model output and will evaluate its performance during the cool season.

As a note for possible future work, winds from the high-resolution ALADIN NWP model can be used as an input. The use of a mesoscale wind model should improve the forecasted/hindcasted waves.

The operational nearshore wave forecasting system, which will be based on the high-resolution wave forecasting model SWAN suitable for the wave propagation in coastal zones, will be created. The high resolution coastal zone wave model SWAN will operate in a one-way connection (nested) to the offshore wave models VAGBUL, WAM or WW3, where the boundary conditions will be retrieved from. The limited area high resolution NWP model ALADIN will provide the coastal zone wave model SWAN with the wind forcing.

As a result of the evaluation of the atmospheric and wave models, improvements to the current operational system of NIMH-BAS will be done. The new system will consist of the best combination of atmospheric and wave models, which will result from the systematic evaluation with in-situ measurements and satellite data of the wind/wave parameters for the Black Sea.

Acknowledgments
This work was carried out in the framework of a collaboration between the Meteorological Service of France Meteo-France and the National Institute of Meteorology and Hydrology Bulgarian Academy of Sciences –NIMH-BAS. We would like to thank Jean-Michel Lefevre for the scientific support.